\documentclass[twocolumn,showpacs,aps,prd,superscriptaddress,floatfix]{revtex4}

\usepackage{axodraw}
\usepackage{graphicx}
\usepackage{dcolumn}
\usepackage{amsmath}
\usepackage{epsfig}
\usepackage{dcolumn}
\usepackage{natbib}
\usepackage{psfrag}

\RequirePackage{xspace}

\usepackage{relsize}

\def\babar{\mbox{\slshape B\kern-0.1em{\smaller A}\kern-0.1em
    B\kern-0.1em{\smaller A\kern-0.2em R}}}

\def\epem       {\ensuremath{e^+e^-}\xspace}
\def\tautau     {\ensuremath{\tau^+\tau^-}\xspace}
\def\ellell     {\ensuremath{\ell^+ \ell^-}\xspace}

\def\nunub      {\ensuremath{\nu{\overline{\nu}}}\xspace}
\def\qqbar {\ensuremath{q\overline q}\xspace}
\def\ccbar {\ensuremath{c\overline c}\xspace}

\def\piz   {\ensuremath{\pi^0}\xspace}
\def\pip   {\ensuremath{\pi^+}\xspace}
\def\pim   {\ensuremath{\pi^-}\xspace}

\def\kaon  {\ensuremath{K}\xspace}

\def\Kp    {\ensuremath{K^+}\xspace}
\def\Km    {\ensuremath{K^-}\xspace}
\def\KS    {\ensuremath{K^0_{\scriptscriptstyle S}}\xspace} 
\def\Kstarz  {\ensuremath{K^{*0}}\xspace}

\def\Kstar   {\ensuremath{K^*}\xspace}

\def\Kstarp  {\ensuremath{K^{*+}}\xspace}

\def\BChtoKChnn {\ensuremath{\Bp \to \Kstarp\nunub}\xspace}

\def\BNeutoKNeunn {\ensuremath{\Bz \to \Kstarz\nunub}\xspace}
\def\BtoKnn {\ensuremath{\B \to \kaon^{*}\nunub}\xspace}

\def\KstarztoKpi {\ensuremath{\Kstarz \to \Kp\pim}\xspace}

\def\KstarptoKspip {\ensuremath{\Kstarp \to \KS\pip}\xspace}
\def\KstarptoKppiz {\ensuremath{\Kstarp \to \Kp\piz}\xspace}

\def\BtoDstlnu   {\ensuremath{B \to D^{(*)} l \nu }\xspace}
\def\cosBY {\ensuremath{\cos\theta_{B,Dl}}\xspace}

\def\Eextra     {\ensuremath{E_{\mbox{\scriptsize{extra}}}}\xspace}
\def\Emiss      {\ensuremath{E_{\mbox{\scriptsize{miss}}}^*}\xspace}
\def\pmiss      {\ensuremath{|\vec{p}_{\mbox{\scriptsize{miss}}}^{\,*}|}\xspace}
\def\cosmiss        {\mbox{$\ctheta_{\mbox{\scriptsize{miss}}}^*$}\xspace}

\def\costhrust      {\mbox{$\ctheta_{\mbox{\scriptsize{\B,T}}}^*$}\xspace}

\def\snunu          {\mbox{$s_{\mbox{\scriptsize{$\nu\nu$}}}$}\xspace}

\def\Bsig    {\ensuremath{B_{\mbox{\scriptsize{sig}}}}\xspace}
\def\Bhad    {\ensuremath{B_{\mbox{\scriptsize{had}}}}\xspace}
\def\Bsl    {\ensuremath{B_{\mbox{\scriptsize{sl}}}}\xspace}

\def\btosnunu   {\ensuremath{b \to s \nunub}\xspace}
\def\btosll     {\ensuremath{b \to s \ellell}\xspace}

\def\Dbar    {\kern 0.2em\overline{\kern -0.2em D}{}\xspace}

\def\Dz      {\ensuremath{D^0}\xspace}
\def\Dzb     {\ensuremath{\Dbar^0}\xspace}
\def\DzDzb   {\ensuremath{\Dz {\kern -0.16em \Dzb}}\xspace}
\def\Dp      {\ensuremath{D^+}\xspace}
\def\Dm      {\ensuremath{D^-}\xspace}
\def\Dpm     {\ensuremath{D^\pm}\xspace}

\def\DpDm    {\ensuremath{\Dp {\kern -0.16em \Dm}}\xspace}
\def\Dstar   {\ensuremath{D^*}\xspace}

\def\Dstarz  {\ensuremath{D^{*0}}\xspace}

\def\Dstarp  {\ensuremath{D^{*+}}\xspace}

\def\Dstarpm {\ensuremath{D^{*\pm}}\xspace}

\def\B       {\ensuremath{B}\xspace}
\def\Bbar    {\kern 0.18em\overline{\kern -0.18em B}{}\xspace}

\def\BB      {\ensuremath{B\Bbar}\xspace} 
\def\Bz      {\ensuremath{B^0}\xspace}
\def\Bzb     {\ensuremath{\Bbar^0}\xspace}
\def\BzBzb   {\ensuremath{\Bz {\kern -0.16em \Bzb}}\xspace}
\def\Bu      {\ensuremath{B^+}\xspace}
\def\Bub     {\ensuremath{B^-}\xspace}
\def\Bp      {\ensuremath{\Bu}\xspace}

\def\BpBm    {\ensuremath{\Bu {\kern -0.16em \Bub}}\xspace}

\def\Y#1S{\ensuremath{\Upsilon{(#1S)}}\xspace}
\def\FourS   {\Y4S}

\def\Z      {\ensuremath{Z^0}\xspace}

\def\BR         {{\ensuremath{\cal B}\xspace}}

\def\ctheta     {\mbox{$\cos\theta$}\xspace}

\def\mes        {\mbox{$m_{\rm ES}$}\xspace}
\def\DeltaE     {\mbox{$\Delta E$}\xspace}

\newcommand{\tev}{\ensuremath{\mathrm{\,Te\kern -0.1em V}}\xspace}
\newcommand{\gev}{\ensuremath{\mathrm{\,Ge\kern -0.1em V}}\xspace}
\newcommand{\mev}{\ensuremath{\mathrm{\,Me\kern -0.1em V}}\xspace}
\newcommand{\kev}{\ensuremath{\mathrm{\,ke\kern -0.1em V}}\xspace}
\newcommand{\ev}{\ensuremath{\mathrm{\,e\kern -0.1em V}}\xspace}
\newcommand{\gevc}{\ensuremath{{\mathrm{\,Ge\kern -0.1em V\!/}c}}\xspace}
\newcommand{\mevc}{\ensuremath{{\mathrm{\,Me\kern -0.1em V\!/}c}}\xspace}
\newcommand{\gevcc}{\ensuremath{{\mathrm{\,Ge\kern -0.1em V\!/}c^2}}\xspace}
\newcommand{\mevcc}{\ensuremath{{\mathrm{\,Me\kern -0.1em V\!/}c^2}}\xspace}

\def\invfb   {\ensuremath{\mbox{\,fb}^{-1}}\xspace}

\def\pep2{PEP-II}

\newcommand{\stat}{\ensuremath{\mathrm{(stat)}}\xspace}
\newcommand{\syst}{\ensuremath{\mathrm{(syst)}}\xspace}

\newcommand{\chisq}{\ensuremath{\chi^2}\xspace}

\newcommand{\BABARPubYear}    {08}
\newcommand{\BABARPubNumber}  {040}

\newcommand{\SLACPubNumber} {13359}
\newcommand{\LANLNumber} {0808.1338}

\def\figurebox#1#2#3{%
    \def\arg{#3}%
    \ifx\arg\empty
    {\hfill\vbox{\hsize#2\hrule\hbox to #2{\vrule\hfill\vbox to #1{\hsize#2\vfill}\vrule}\hrule}\hfill}%
    \else
    {\hfill\epsfbox{#3}\hfill}%
    \fi}

\begin{document}

\preprint{\babar-PUB-\BABARPubYear/\BABARPubNumber} 
\preprint{SLAC-PUB-\SLACPubNumber} 

\begin{flushleft}
\babar-PUB-\BABARPubYear/\BABARPubNumber\\
SLAC-PUB-\SLACPubNumber\\
arXiv:\LANLNumber\ [hep-ex]\\[10mm]
\end{flushleft}

\title{{\large \bf Search for \BtoKnn decays} }

\date{\today}

\begin{abstract}
We present a search for the decays \BtoKnn using 454$\times10^{6}$ \BB pairs
collected at the \FourS resonance with the \babar\ detector at the SLAC PEP-II
\B-Factory. We first select an event sample where one \B is reconstructed in a
semileptonic or hadronic mode with one charmed meson. The remaining particles in
the event are then examined to search for a \BtoKnn decay. The charged \Kstar is
reconstructed as $\Kstarp \to \KS \pip$ or $\Kstarp \to \Kp \piz$; the neutral \Kstar is
identified in \KstarztoKpi mode. We establish upper limits at $90\%$ confidence
level of $\BR(\BChtoKChnn) < 8 \times 10^{-5}$, $\BR(\BNeutoKNeunn) < 12 \times
10^{-5}$, and $\BR(\B \to \Kstar \nunub) < 8 \times 10^{-5}$.
\end{abstract}


\author{B.~Aubert}
\author{M.~Bona}
\author{Y.~Karyotakis}
\author{J.~P.~Lees}
\author{V.~Poireau}
\author{E.~Prencipe}
\author{X.~Prudent}
\author{V.~Tisserand}
\affiliation{Laboratoire de Physique des Particules, IN2P3/CNRS et Universit\'e
de Savoie, F-74941 Annecy-Le-Vieux, France }
\author{J.~Garra~Tico}
\author{E.~Grauges}
\affiliation{Universitat de Barcelona, Facultat de Fisica, Departament ECM,
E-08028 Barcelona, Spain }
% ...
\collaboration{The \babar\ Collaboration}
\noaffiliation

\maketitle

\begin{figure}[b]
\includegraphics[width=6cm]{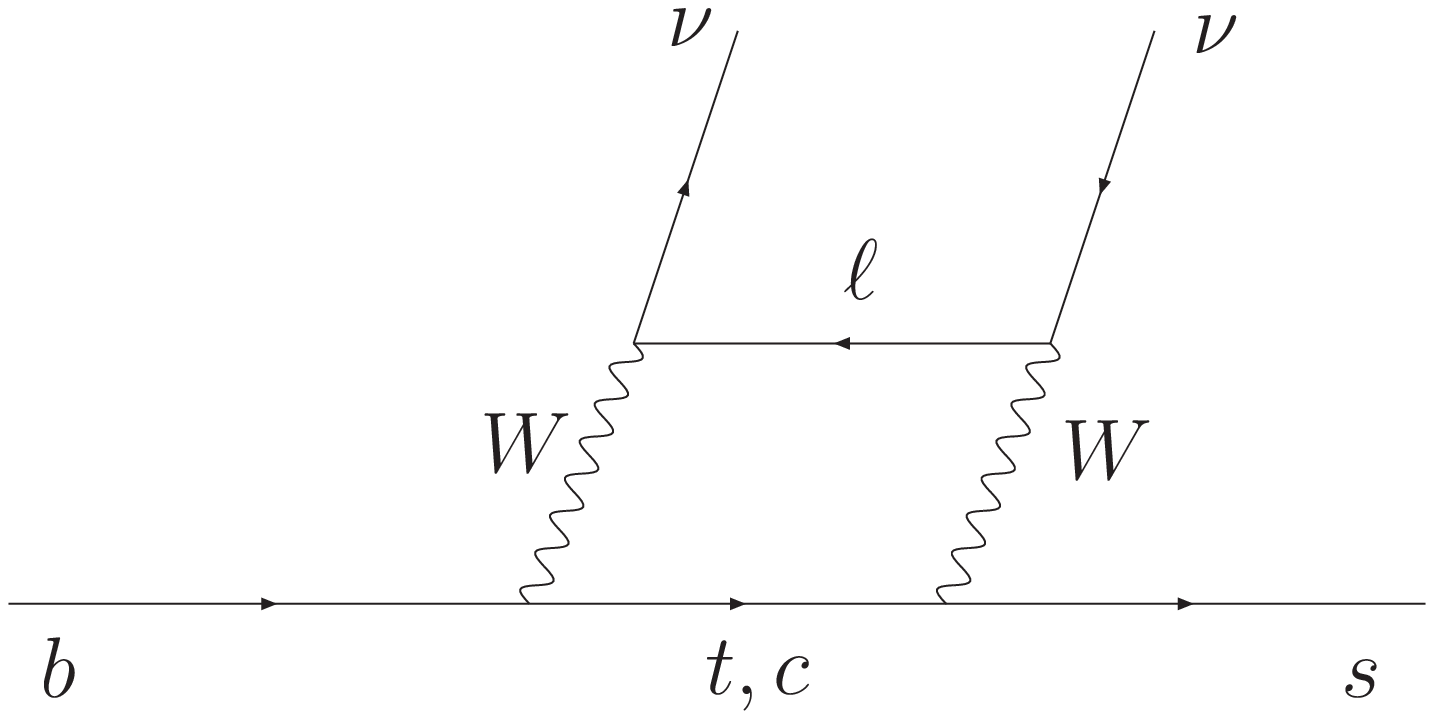}\\
\vspace{0.5cm}
\includegraphics[width=6cm]{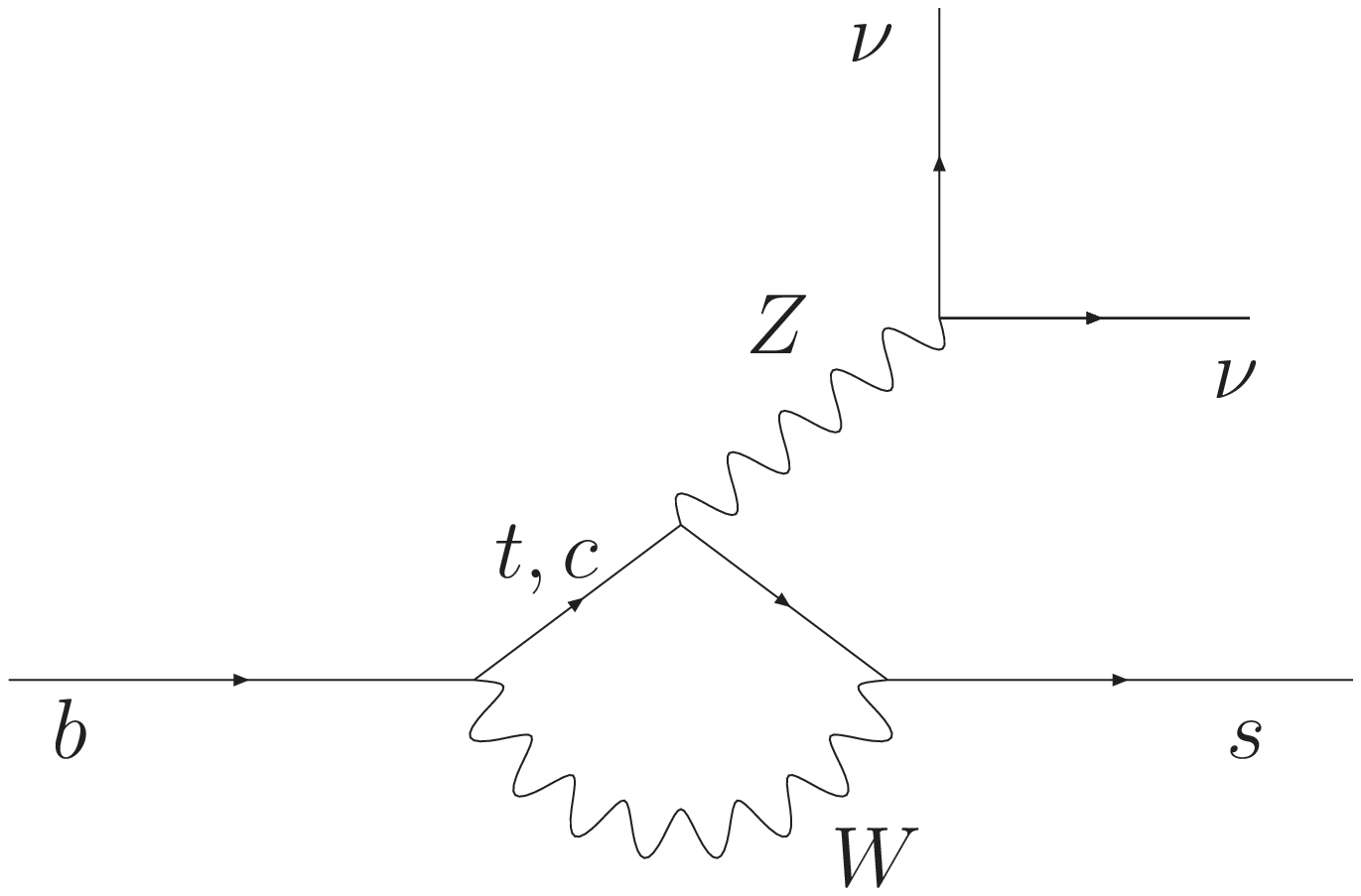}
\caption{SM diagrams for $b \to s \nu \overline \nu$ transitions.}
\label{fig:feyn}
\end{figure}

In the Standard Model (SM) the \btosnunu process occurs via
one-loop box or electroweak penguin diagrams, as shown in Fig.~\ref{fig:feyn},
and it is therefore expected to be highly suppressed. Due to the absence of
photon penguin contributions and long distance effects, the corresponding rate
is predicted in the SM with smaller theoretical uncertainties than \btosll.
In particular, the SM branching fraction for $\B \to \Kstar \nunub$ is expected
to be $(1.3^{+0.4}_{-0.3}) \times 10^{-5}$~\cite{BHI}. However, 
this could be enhanced in many new physics (NP) scenarios, where several
mechanisms contribute to the rate. In
Ref.~\cite{BHI} non-standard \Z coupling contributions are computed, giving an
enhancement of up to a factor 10. Moreover, new sources of missing energy, such as 
light dark matter candidates~\cite{bird} or unparticles~\cite{georgi,aliev},
could contribute to the rate and produce a final state with a
$\Kstar$~\cite{cc} plus missing energy. The kinematics of the decay is described
in terms of $\snunu = m^2_{\nu\nu}/m^2_{\B}$, where $m_{\nu\nu}$ is the
invariant mass of the neutrino pair and $m_{\B}$ is the \B
meson mass. NP effects can strongly affect the shape of the \snunu
distribution~\cite{BHI,aliev}, and this is taken into account in the present
work to obtain a model independent limit.

A previous search by the Belle Collaboration~\cite{belle} sets
upper limits of $\BR(\BChtoKChnn)< 1.4 \times 10^{-4}$ and $\BR(\BNeutoKNeunn)<
3.4 \times 10^{-4}$ at 90\% confidence level~\cite{cc}.

In this paper we present the first \babar\ search for both neutral and charged
\BtoKnn decays. The analysis is
based on the data collected with the \babar\ detector~\cite{NIM}
at the PEP-II storage ring. The sample corresponds to an integrated luminosity
of 413~\invfb at the \FourS resonance, consisting of about $454 \times 10^6$ \BB
pairs. An additional sample of $41\,\invfb$ was collected at a
center of mass energy $40\,\mev$ below the \FourS resonance in order to study
continuum events: $\epem \to \qqbar$ ($q$ = $u,d,s,c$) and $\epem \to \tautau$. 
Charged-track reconstruction is provided by a silicon vertex detector
and a drift chamber operating in a $1.5\,T$ magnetic field. Particle
identification is based on the energy loss in the tracking system and the
Cherenkov angle in an internally reflecting ring-imaging Cherenkov detector.
Photon detection is provided by a CsI(Tl) electromagnetic calorimeter (EMC).
Finally, muons are identified by the instrumented magnetic-flux return. 

Pairs of photons with invariant mass between 115 and $150\,\mevcc$ are 
considered as \piz candidates.
The \KS candidates are reconstructed from pairs of oppositely charged pions.

A GEANT4-based~\cite{geant} Monte Carlo (MC) simulation is used to model the
detector response and test the analysis technique. Approximately $13 \times
10^6$ events are simulated where one $B$ meson decays to a signal candidate
mode and the other $B$ decay is unconstrained (signal MC sample), and the
kinematics of the signal decay is described by a pure phase space model.
Simulated generic \BB and continuum samples are used to investigate the
background contamination and perform systematic studies.

Due to the presence of two undetected neutrinos in the final state, the
\BtoKnn decays cannot be fully reconstructed. Hence, one of the two
$B$ mesons produced in the \FourS decay (the \emph{tagging} \B) is reconstructed
in a semileptonic (\Bsl) or a hadronic (\Bhad) mode containing a charmed meson.
Then a $\Kstar$ and missing energy are searched for in the rest of the event
(ROE), defined as the set of tracks and EMC clusters not associated with the
tagging \B. The two tagging strategies provide non overlapping samples and the
corresponding results can be combined as independent measurements. Selection
criteria are applied to suppress the background contamination and an extended
maximum likelihood 
fit is performed to extract the signal yields ($N_s$), which are finally used
to determine the decay branching fractions ($\mathcal{B}$). In general, these
can be written as:
\begin{equation}
 \BR = \frac{N_s}{\varepsilon \cdot N_{B\bar{B}}} \, ,
\label{eq:BR}
\end{equation}
where $\varepsilon$ is the total signal efficiency measured with the 
signal MC sample and
$N_{B\bar{B}}$ the number of produced \BB pairs.
In the semileptonic (SL) tagged analysis we
adopt Eq.~(\ref{eq:BR}) and use control samples to correct for small data/MC
disagreements in the efficiency.
In the hadronic (HAD) analysis, in order to avoid large systematic uncertainties
associated with the MC estimate of the reconstruction efficiency for the
\Bhad, we normalize the branching fraction with respect to the number of data
events with a correctly reconstructed \Bhad ($N_{\Bhad}$):
\begin{equation}
 \BR = \frac{N_s}{\varepsilon_{\mbox{\scriptsize{\Bsig}}} \cdot
N_{\mbox{\scriptsize{\Bhad}}}} \cdot
\frac{\varepsilon^{\mbox{\scriptsize{\BB}}}_{\mbox{\scriptsize{\Bhad}}}}{
\varepsilon^{\mbox{\scriptsize{\Kstar\nunub}}}_{\mbox{\scriptsize{\Bhad}}}} \, ,
\label{eq:BRhad}
\end{equation}
where $\varepsilon_{\Bsig}$ is the efficiency related to the signal side
reconstruction and selection, while $\varepsilon^{\BB}_{\Bhad}$ and
$\varepsilon^{\Kstar\nunub}_{\Bhad}$ are the \Bhad reconstruction efficiencies
in events with generic \BB decays and events containing the signal 
process, respectively; to account for differences among them, observed in the
MC samples, their
ratio $\varepsilon^{\mbox{\scriptsize{\Kstar\nunub}}}_{\mbox{\scriptsize{
\Bhad}}}/\varepsilon^{\mbox{\scriptsize{\BB}}}_{\mbox{\scriptsize{\Bhad}}}$ is
used in Eq.~\ref{eq:BRhad} as a correction factor.

\begin{figure}[htb!]
\includegraphics[width=4.3cm]{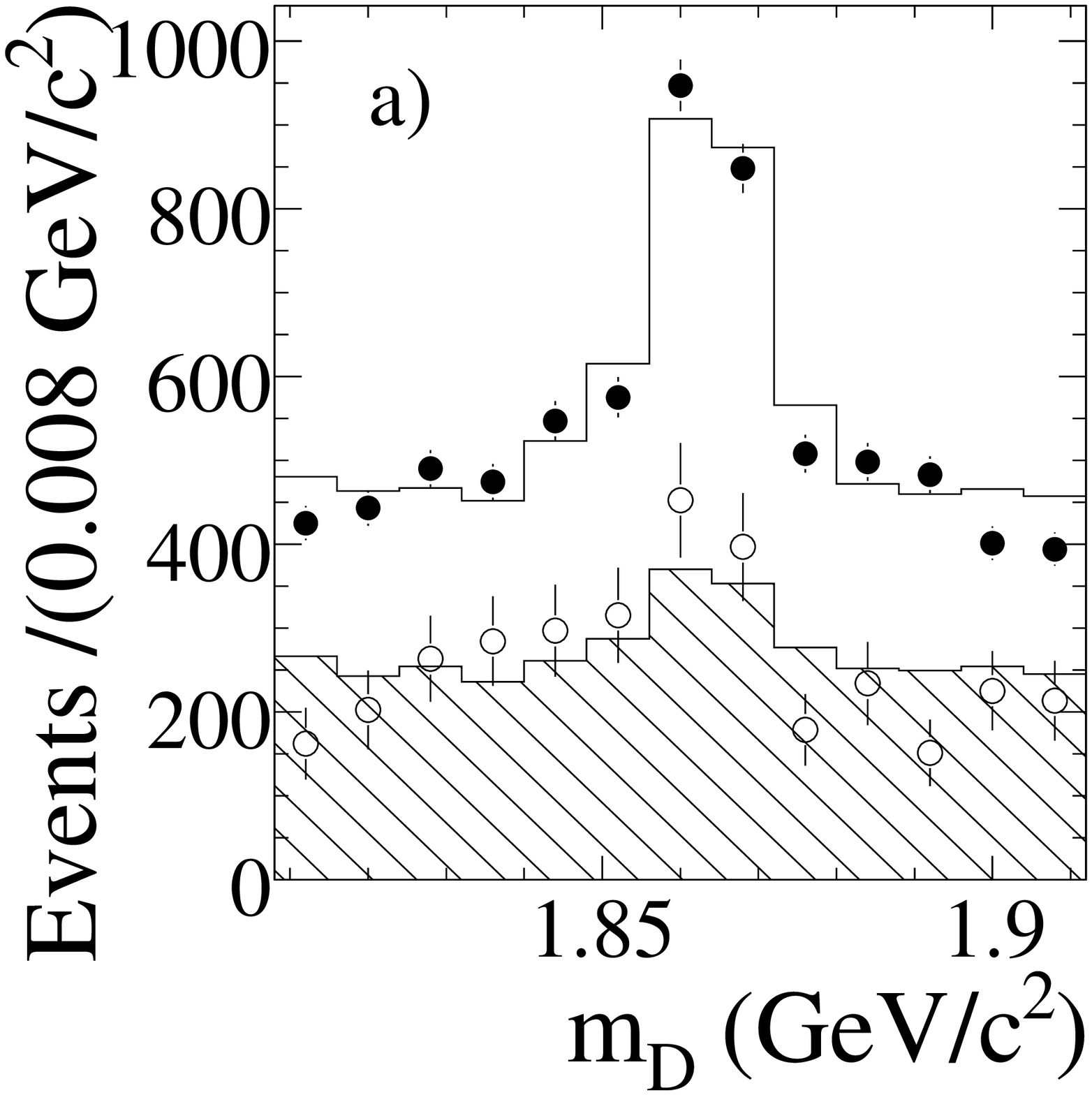}
\hspace{-0.3cm}
\includegraphics[width=4.3cm]{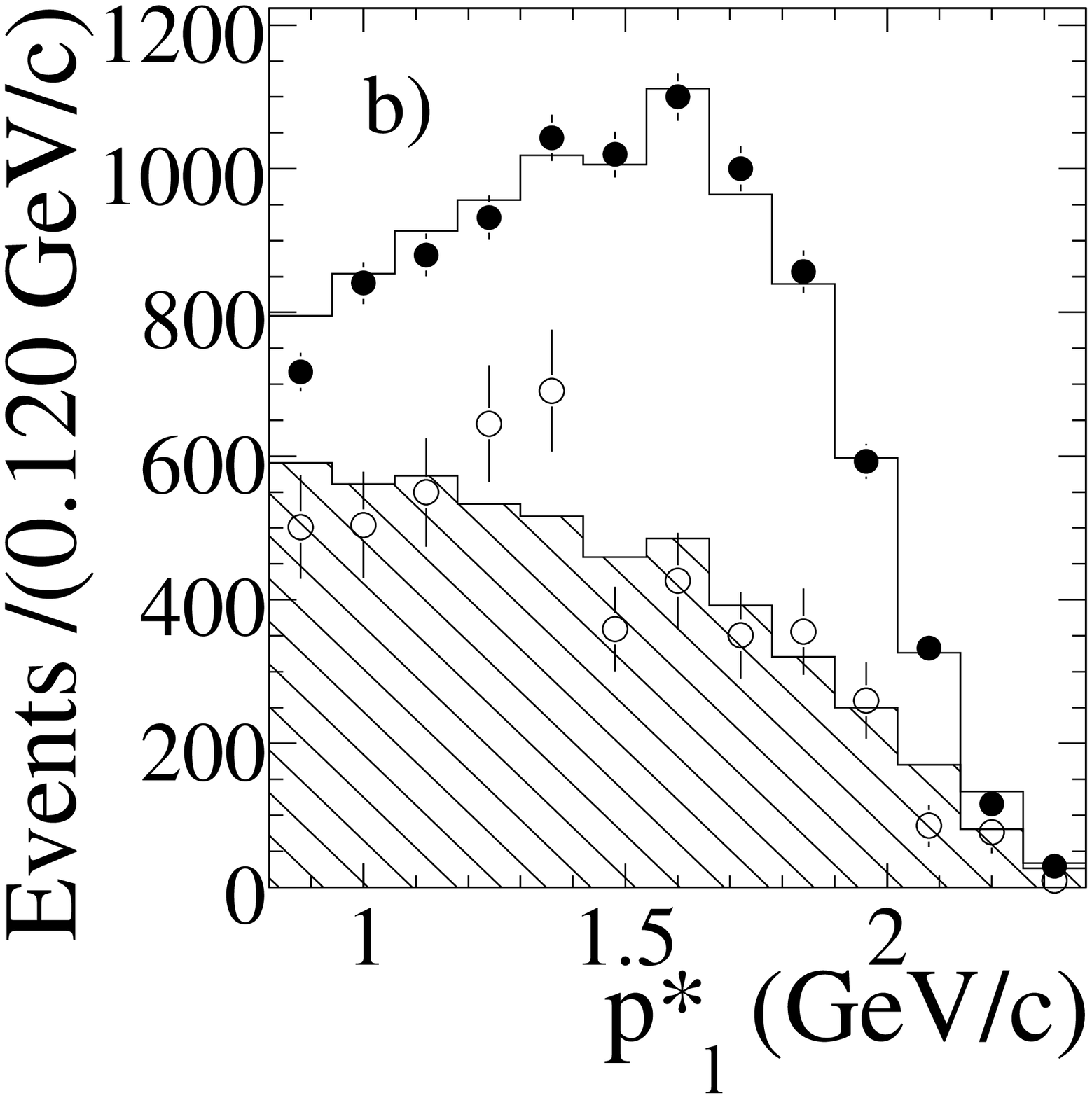}
\caption{The $D$ mass a) and the CM momentum of the \Bsl lepton b),
in the SL analysis from MC simulations (histogram, the hatched area shows the
continuum contribution), on-peak data ($\bullet$) and luminosity scaled off-peak
data ($\circ$). Events where in the signal side a \KstarptoKppiz channel is
reconstructed are shown.}
\label{fig:sltag}
\end{figure}

The event selection starts from the reconstruction of the tagging \B. In the
SL analysis, we search for a \BtoDstlnu decay.
Neutral $D$ mesons are reconstructed in the  $\Km\pip$,
$\Km\pip\piz$, $\Km\pip\pip\pim$ and $\KS\pip\pim$ modes.
Charged $D$ mesons are reconstructed in the $\Km\pip\pip$ and $\KS\pip$
final states. The \Dstarz candidates are reconstructed in the $\Dstarz \to \Dz
\gamma$ channel and the \Dstarp candidates in the $\Dstarp \to
\Dz \pip$ or $\Dstarp \to \Dp \piz$ channels. Finally, a lepton (electron or
muon) candidate is associated to the $D$ meson and a kinematical fit is
performed to find the \Bsl decay vertex.
Preliminary selection requirements are applied on the $D$ mass 
(within $0.07\,\gevcc$ of the nominal mass in the $\Km\pip\piz$ mode, within
$0.04\,\gevcc$ elsewhere) and the momentum of the lepton in the center of mass
(CM) frame ($|\mathbf p^{\,*}_l| > 0.8\,\gevc$). We also require the CM angle
between the \Bsl and the $D^{(*)}l$ pair to satisfy $-5.0 < \cosBY < 1.5$,
where \cosBY can be calculated from the $D^{(*)}l$ four-momentum assuming that
only one massless particle is missing:
\begin{equation}
\label{eq:cosBY}
 \cosBY = \frac{2 E^*_{B,exp} E^*_{Dl} - m^2_B - m^2_{Dl}}{2|
\mathbf{p}^{\,*}_{B,exp}||\mathbf{p}^{\,*}_{Dl}|} \, .
\end{equation}
In Eq.~\ref{eq:cosBY}, $m_B$ is the nominal $B$ mass, 
$E^*_{B,exp}$ and $|\mathbf p^{\,*}_{B,exp}|$ are the expected $B$ energy and momentum, fixed by the
energies of the beams and evaluated in the CM frame, and $|\mathbf
p^{\,*}_{Dl}|$
is the $D^{(*)}l$ pair momentum in the CM frame. Values of \cosBY out of the
physical range [-1,1] are due to resolution effects and missing particles in the
$Dl$ reconstruction. The distributions of the $D$ mass and the lepton momentum in
the CM frame, after the reconstruction of the signal \B, are shown in
Fig.~\ref{fig:sltag}. 
The plots are made after the signal reconstruction
since in case of multiple \Bsl candidates, 
the selection of the best one depends on the signal side reconstruction
too, as will be discussed later; events where in the signal side a
\KstarptoKppiz channel is reconstructed are shown.
If one \BtoDstlnu candidate can be reconstructed in the ROE with the same
procedure adopted for the tag side, the event is selected as a control sample of
\emph{double-tagged events} for systematic studies. 

\begin{figure}[htb!]
\includegraphics[width=4.35cm]{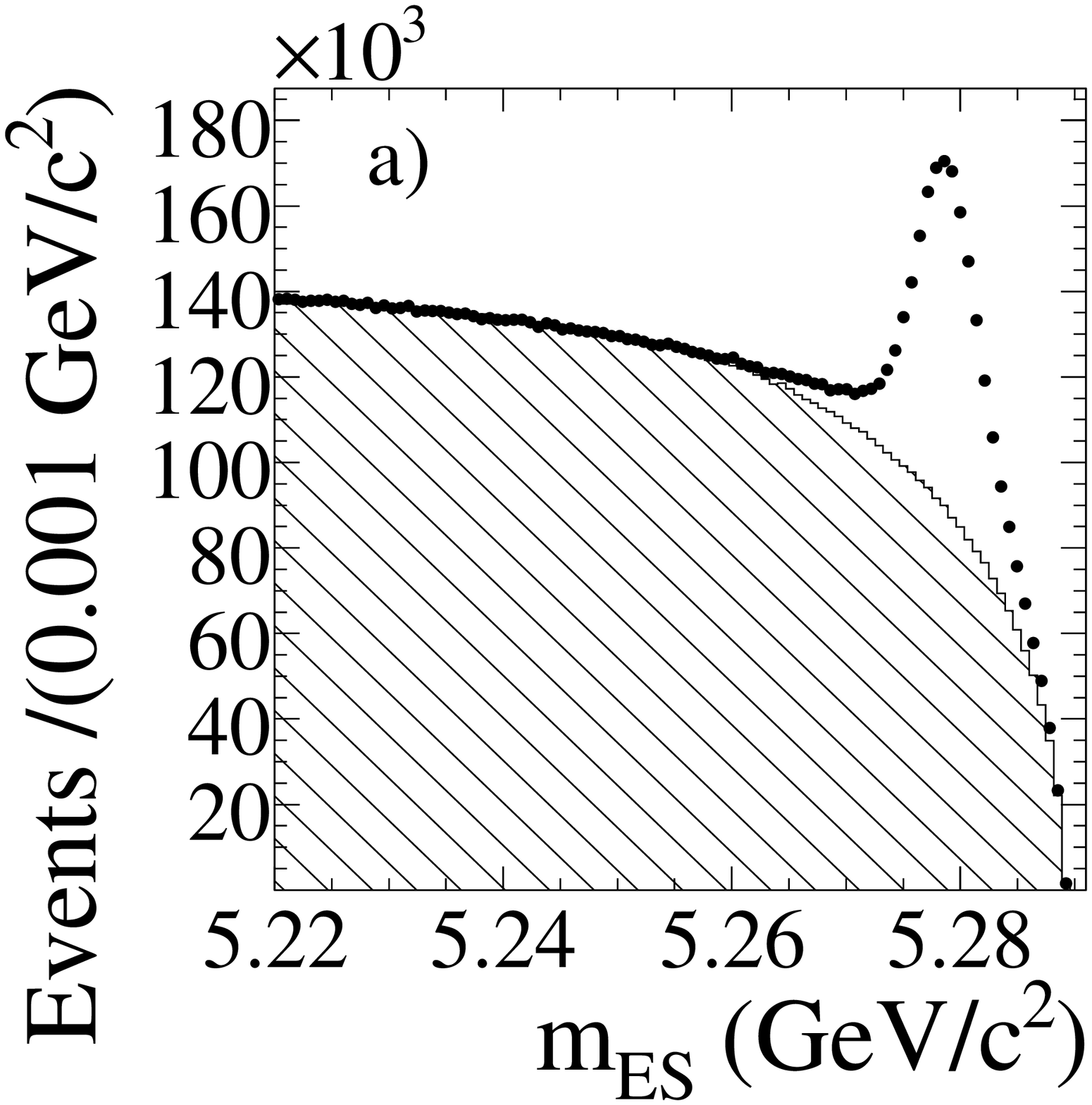}
\hspace{-0.3cm}
\includegraphics[width=4.35cm]{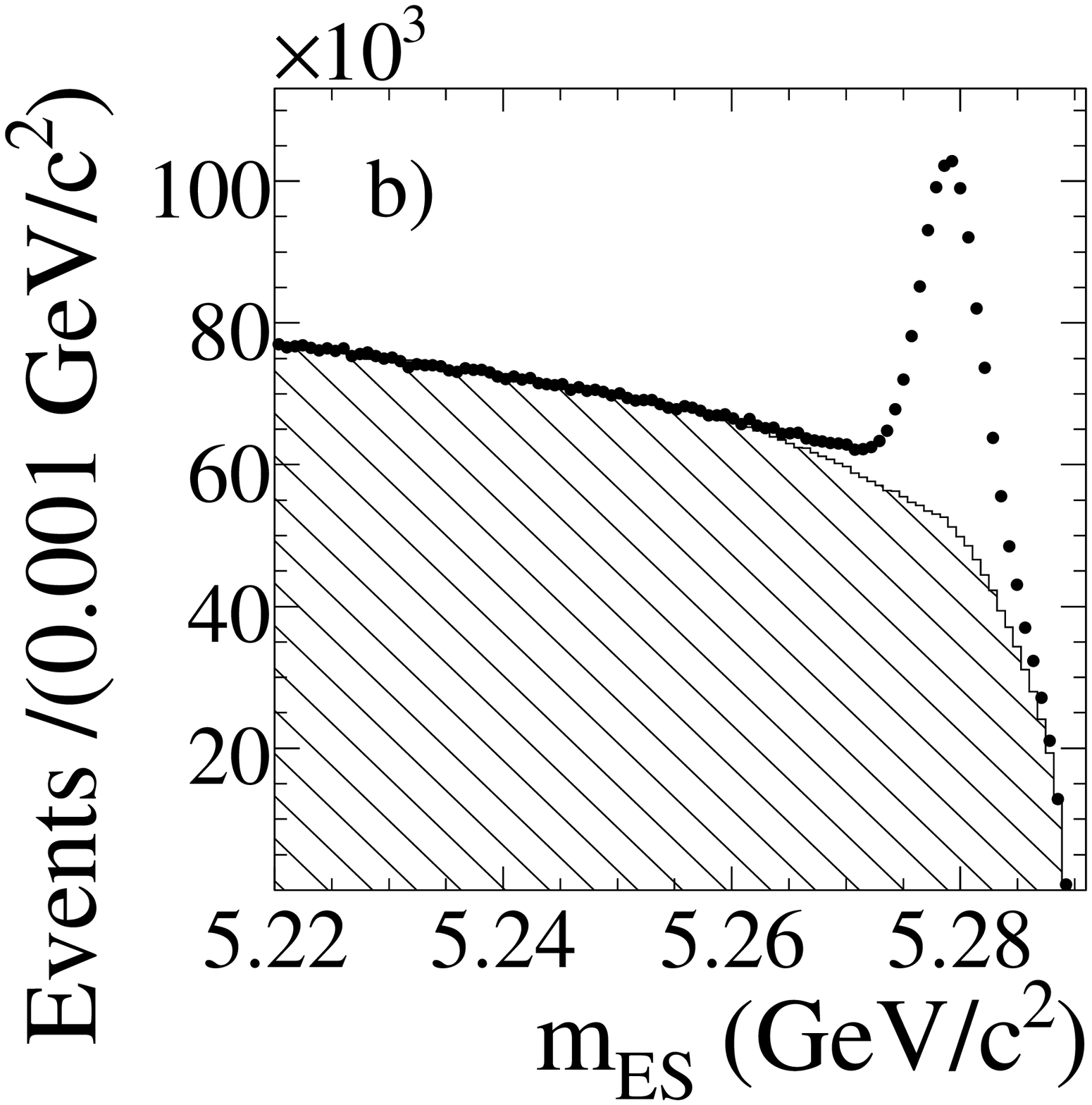}
\vspace{-0.6cm}
\caption{The \mes distributions for charged a) and neutral b) \Bhad.
The points represent the on-peak data and the hatched area shows the
estimated background contribution.}
\label{fig:hadtag}
\end{figure}

In the HAD analysis, we reconstruct \Bhad decays of the type $\Bbar \rightarrow
D
Y$, where $D$ refers to a charm meson, and $Y$ represents a collection
of hadrons with a total charge of $\pm 1$, composed of $n_1\pi^{\pm}+n_2
K^{\pm}+n_3 \KS+n_4\piz$, where $n_1+n_2 \leq 5$, $n_3 \leq 2$, and
$n_4 \leq 2$. Using $\Dz(D^+)$ and $D^{*0}(D^{*+})$ as seeds for
$B^-(\Bzb)$ decays, we reconstruct about 1000 different decay chains.
Charmed mesons are reconstructed in the same final states
used in the SL analysis, along with the additional channels $\Dp \to \Kp \pip
\pim \piz$, $\KS \pip \pip \pim$, $\KS \pip \piz$ and $\Dstarz \to \Dz \piz$.
\Bhad candidates are selected by the two kinematical variables:
\begin{eqnarray}\nonumber
\mes & = &
\sqrt{E^{*2}_{\mbox{\scriptsize{beam}}}-|\mathbf p^{\,*}_{B}|^2}\\
\DeltaE & = & E^*_{\B}-E^*_{\mbox{\scriptsize{beam}}} \, ,
\label{eq:tagDef}
\end{eqnarray}
where $E^*_{\mbox{\scriptsize{beam}}}$ is the beam energy and
$E^{*}_{\mbox{\scriptsize{\B}}}$ and 
$\mathbf p^{\,*}_{\mbox{\scriptsize{\B}}}$ are
the energy and the momentum of the \Bhad in the CM frame.
For correctly tagged \B candidates, the \mes distribution peaks at the nominal
\B mass value and \DeltaE at zero. Hence, a selection is applied by requiring
$-0.09 < \DeltaE < 0.05\,\gev$ and $5.270 < \mes < 5.288\,\gevcc$. 
The number of correctly reconstructed \Bhad events, to be used in 
Eq.~(\ref{eq:BRhad}), is extracted from the \mes distribution of on-peak data.
Background events are classified in four categories:
combinatorial \BzBzb, combinatorial \BpBm, $\epem \to \ccbar$ and $\epem \to
\qqbar$ ($q = u,d,s$). Other sources of background are found to be
negligible. For each category, we extract the \mes shape from MC simulations.
The normalizations of the continuum contributions are taken from off-resonance
data, scaled by the luminosity. The normalization of the \BB contribution is
extracted from a \chisq fit in the $5.22 < \mes < 5.26\,\gevcc$ region. The
number of misreconstructed \Bhad in the signal region is extrapolated from the
fit and subtracted from the data yield. In Fig.~\ref{fig:hadtag} 
the \mes distributions for charged and neutral \Bhad are shown:
the on-peak data are superimposed to the estimated background 
contribution.
After background subtraction, including correction factors and
systematic uncertainties that will be discussed later, 
we determine $N_{\Bhad} = (7.175 \pm 0.008\stat \pm 0.222\syst) \times 10^5$
for neutral \Bhad and $N_{\Bhad}=(10.128 \pm 0.010\stat \pm 0.344\syst)\times
10^5$
for the charged \Bhad.

For each reconstructed tagging \B, we search for a \Kstar
candidate in the ROE. A neutral \Kstar can be reconstructed in the $\Kp\pim$
mode, while a charged \Kstar can be reconstructed in the $\KS\pip$
and $\Kp\piz$ channels. The number of tracks in the ROE is required to match
exactly the number of expected tracks for the selected mode. The signal \B must
have opposite flavor (inferred from the \Kstar flavor) with respect to the
tagging \B.

If more than one $\Bsl (\Bhad)\,-\,\Bsig$ pair has been reconstructed, only one
of them is selected. In the SL analysis, we adopt a Bayesian approach to define
the probability that both signal and tag side have been correctly
reconstructed, given a set {\bf x} of observed quantities:
\begin{equation}
 P(TT|\mathbf{x}) = \frac{P(\mathbf{x}|TT)P(TT)}{\sum_i
 P(\mathbf{x}|i)P(i)}~,~~~i = \mbox{$TT$, $TF$, $FT$, $FF$}
\label{eq:best}
\end{equation}
where $TT$ ($FF$) indicates that both sides are correctly (wrongly)
reconstructed
and $TF$ ($FT$) that only the tag (signal) side is correctly reconstructed.
The candidate with the highest
$P(TT|\mathbf{x})$ is retained. The set {\bf x} is composed of the \chisq
probabilities of the \Bsl and the \Kstar vertex fit.
The corresponding likelihoods and prior probabilities are modeled from MC
simulations with truth information to identify the correctly
reconstructed candidates. In the HAD analysis, if more than one \Bhad is
reconstructed, the best one is selected according to the smallest \DeltaE;
if there are multiple \Kstar candidates associated to the best \Bhad, the
one with a reconstructed mass closest to the world average value~\cite{PDG} is
chosen.
\begin{figure*}[t!]
\includegraphics[width=5cm]{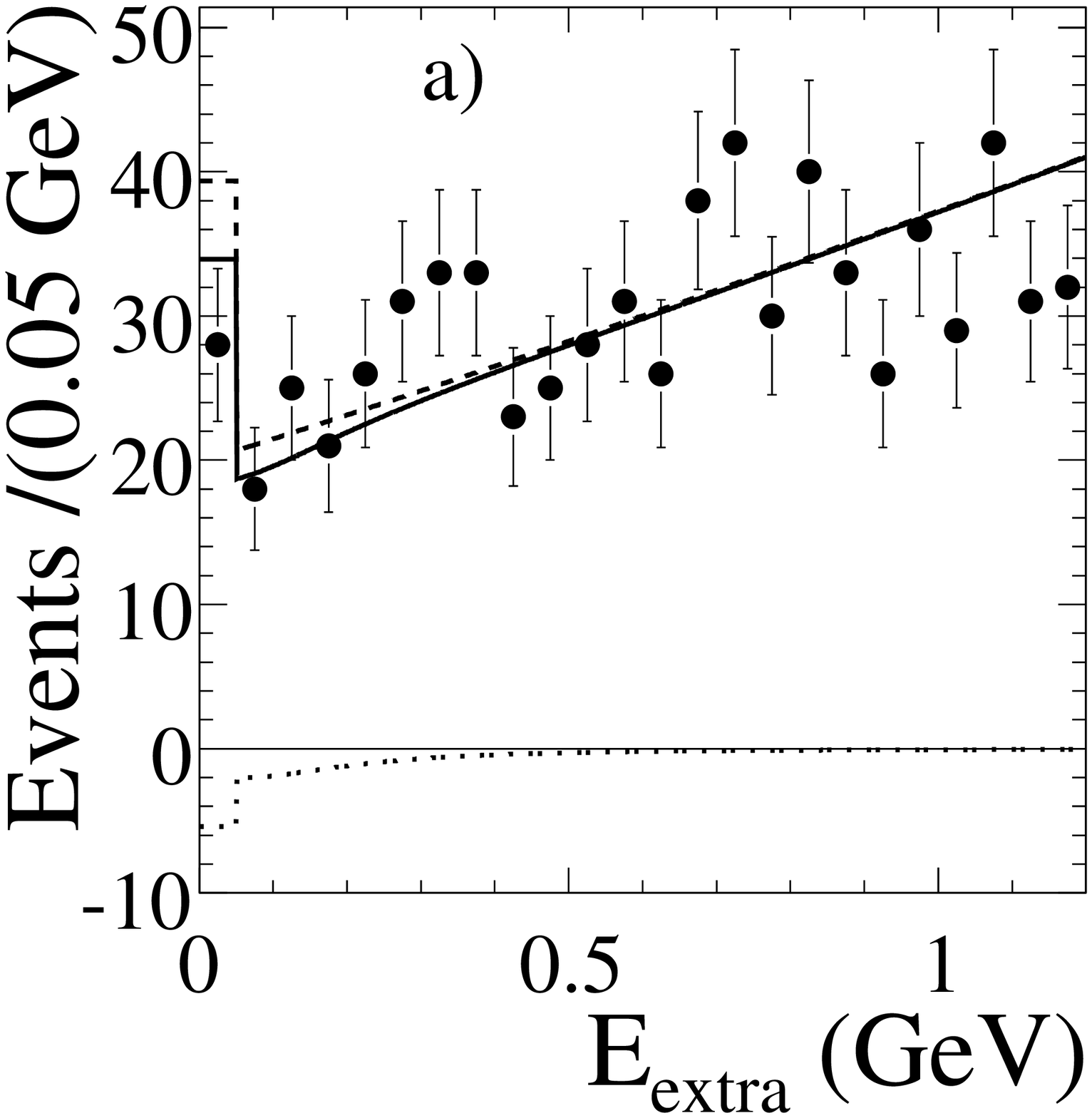}
\includegraphics[width=5cm]{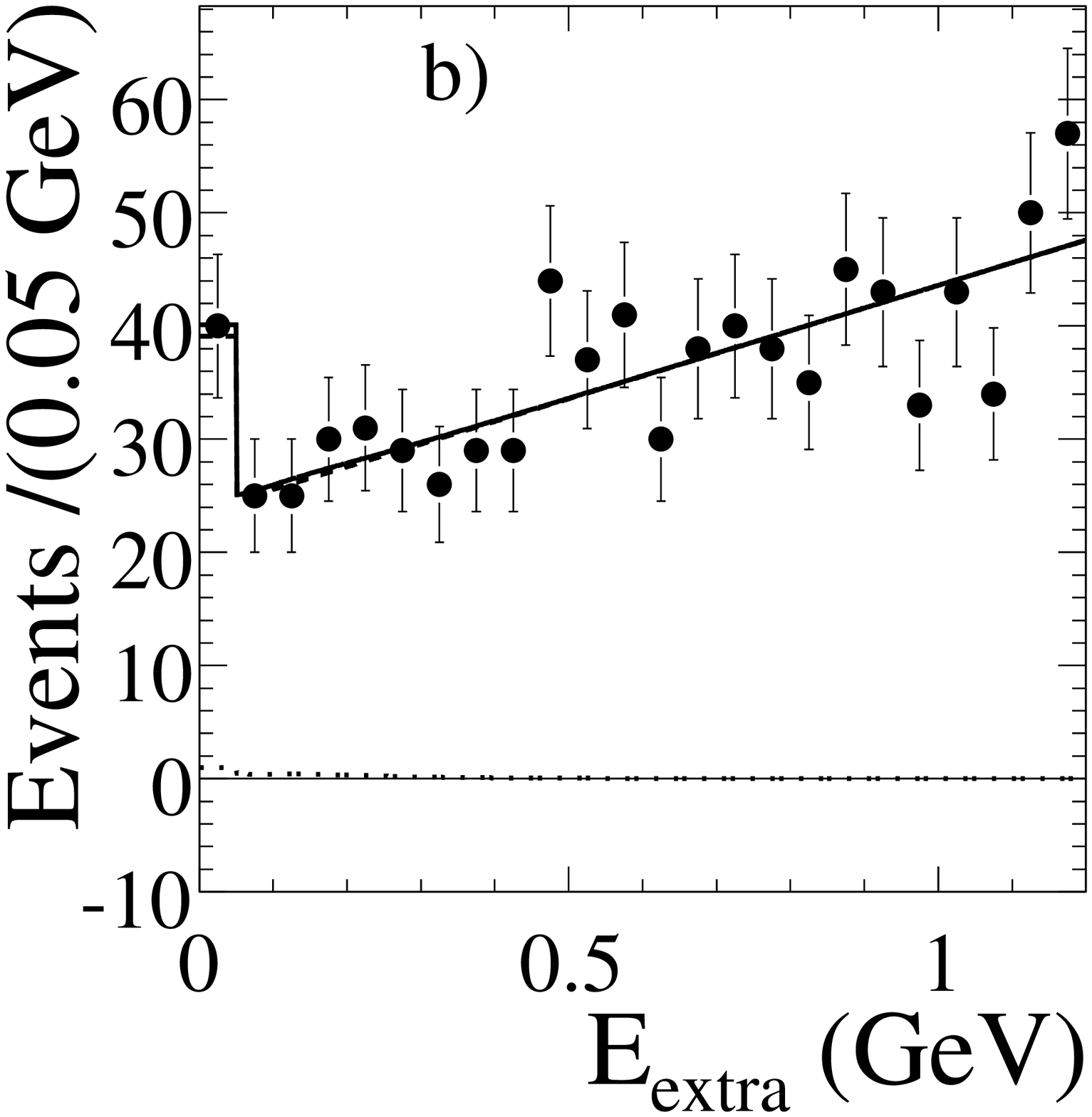}
\includegraphics[width=5cm]{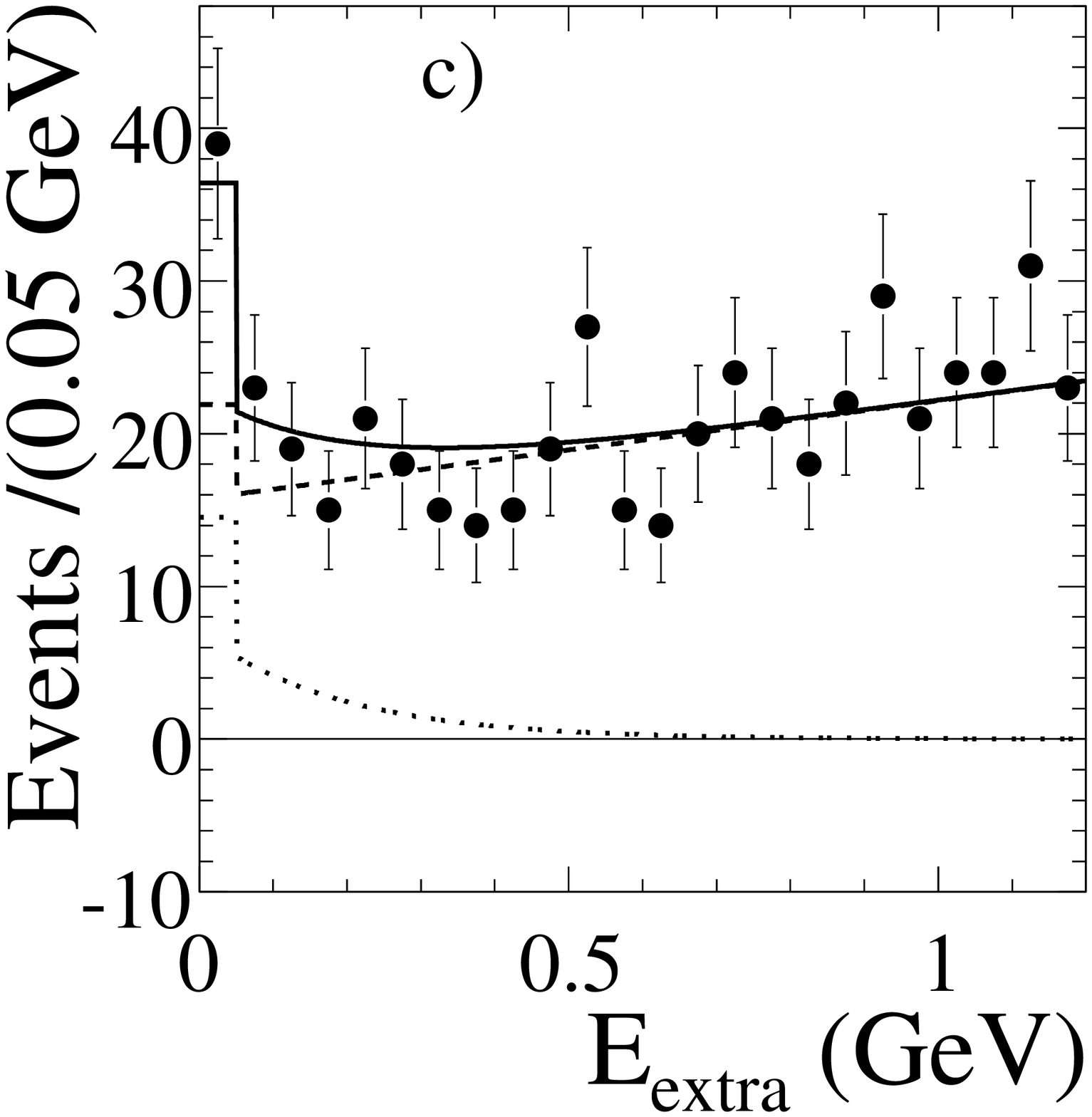}\\
\includegraphics[width=5cm]{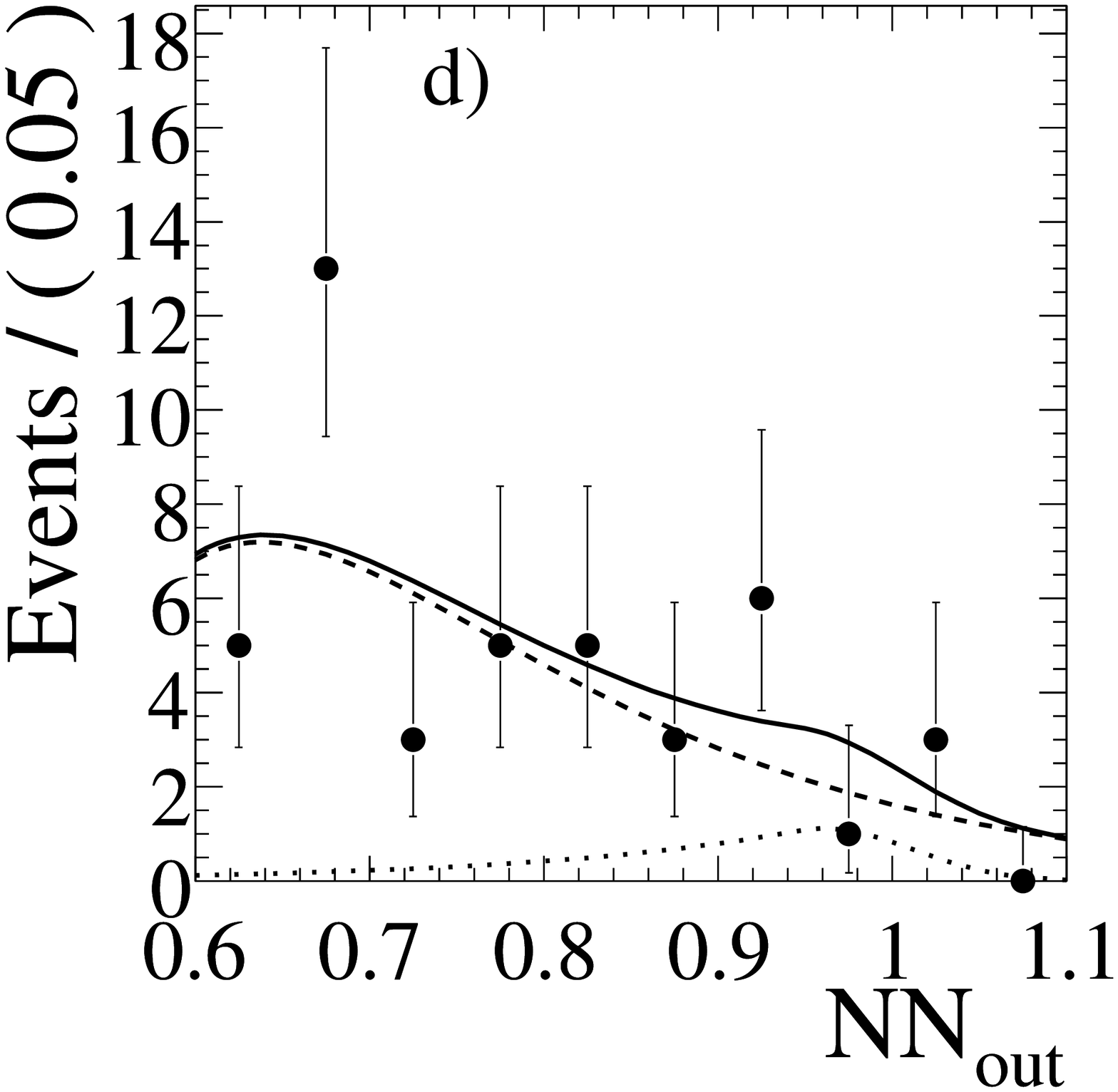}
\includegraphics[width=5cm]{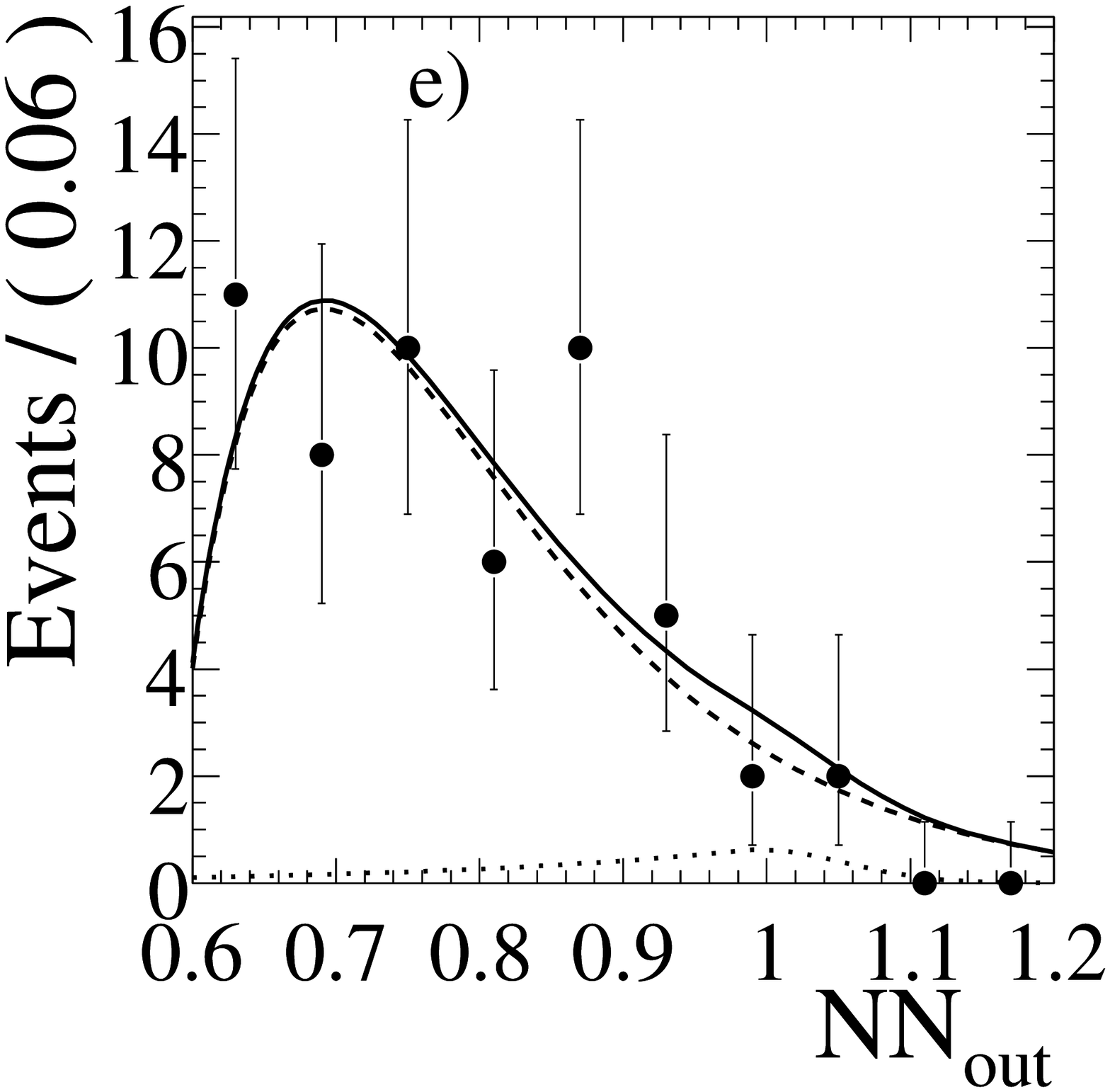}
\includegraphics[width=5cm]{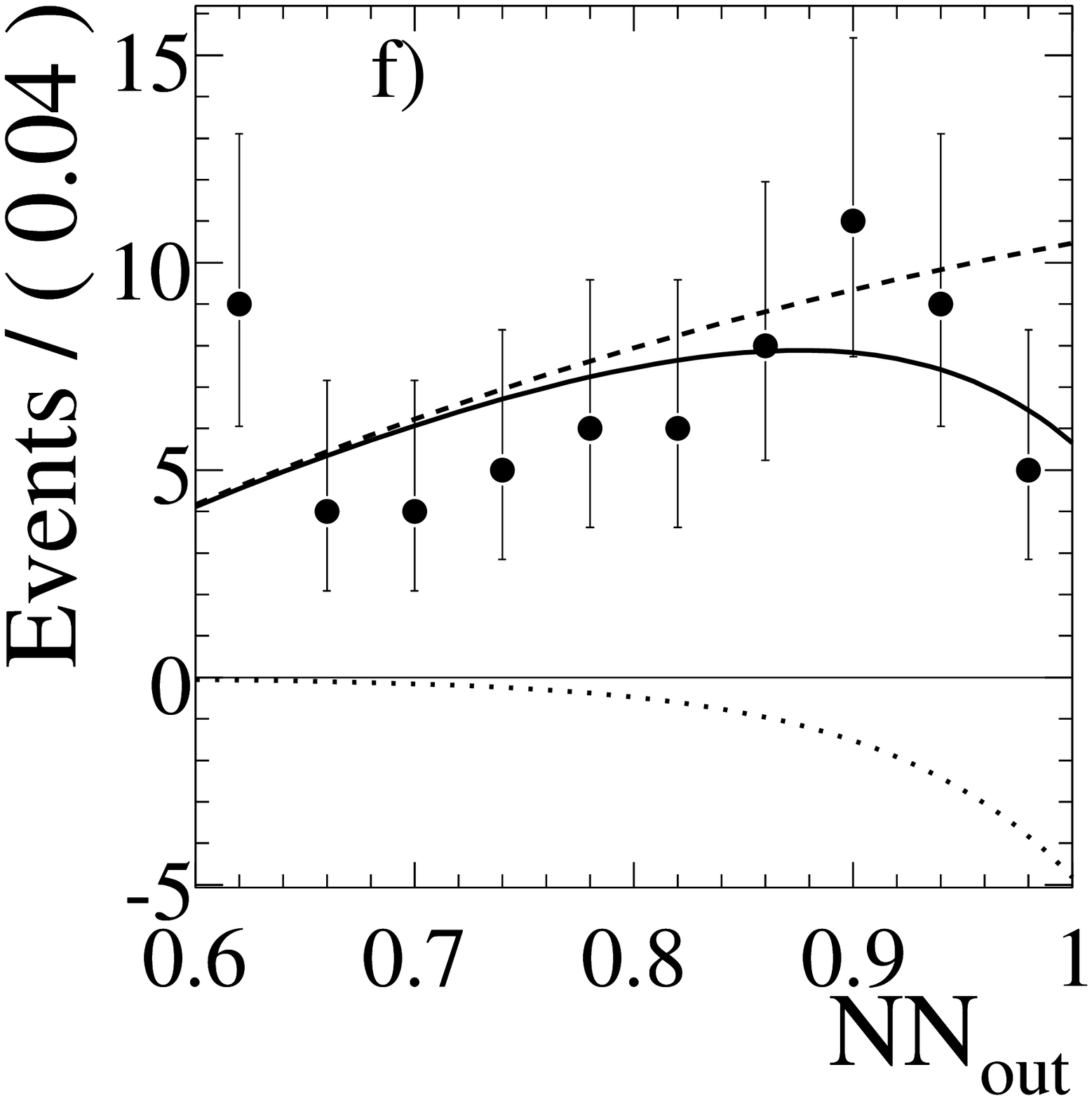}
\caption{Fit results for the extra EMC energy \Eextra a)~-~c) for the SL
analysis and the neural network output $NN_{\mbox{\scriptsize{out}}}$ d)~-~f)
for the HAD analysis. From left to right, \KstarptoKppiz, \KstarptoKspip and
\KstarztoKpi. Data are shown as points, and the fit result is shown with a solid
line. The dotted and dashed lines show the estimated signal and background
contributions respectively.}
\label{fig:fit}
\end{figure*} 

\begin{table}[t!hb]
\begin{center}
\caption{Discriminating variables used in SL and HAD analyses and specific
selection requirements. Values given in the squared brackets represent
the lower and upper selection criteria imposed on the respective quantity.}
\begin{tabular}{lccc}
\hline
\hline
 Variable & Mode & \multicolumn{2}{c}{Range} \\
           &      & \multicolumn{2}{c}{SL~~~~~~~~~~~HAD}      \\    
\hline
\costhrust  & \KstarptoKppiz  & [-0.98,0.97] &    \\
	    & \KstarptoKspip  & [-0.99,1.00] & [-0.95,0.95] \\
            & \KstarztoKpi    & [-1.00,1.00] &    \\
\hline
$R_2$       & \KstarptoKppiz  & [0.01,0.82] & \\
	    & \KstarptoKspip  & [0.01,0.71] & [0.00,0.70]\\
            & \KstarztoKpi    & [0.00,0.80] & \\
\hline
$m_{\Kstar}$  & \KstarptoKppiz  & [0.83,0.97] & \\
(\gevcc)      & \KstarptoKspip  & [0.85,0.95] & [0.84,0.96]\\
              & \KstarztoKpi    & [0.84,0.97] & \\
\hline
$m_{\KS}$   & \KstarptoKppiz     & -- & -- \\
(\gevcc)    & \KstarptoKspip     & [0.49,0.50] & [0.49,0.51]\\
            & \KstarztoKpi       & -- & -- \\
\hline
$\Emiss + \pmiss$ & \KstarptoKppiz  & [5.81,8.82] & \\
(\gev)	             & \KstarptoKspip  & [5.01,8.73]  & -- \\
                     & \KstarztoKpi    & [5.11,9.01] & \\
\hline
\cosmiss  & \KstarptoKppiz  & [-0.90,0.88] & \\
	  & \KstarptoKspip  & [-0.88,0.85] & [-0.90,0.90]\\
          & \KstarztoKpi    & [-0.95,0.89] & \\
\hline
\Eextra   & \KstarptoKppiz  &  & \\
(\gev)    & \KstarptoKspip  & [0.00,1.20] & -- \\
          & \KstarztoKpi    &  & \\
\hline
$|\mathbf p^{\,*}_l|$ & \KstarptoKppiz  & [0.95,2.40] & \\
(\gevc)	              & \KstarptoKspip  & [0.80,2.40] & -- \\
                      & \KstarztoKpi    & [0.84,2.48] & \\
\hline
\hline
\end{tabular}
\label{tab:selection}
\end{center}
\end{table}

\begin{table}[b!]
\begin{center}
\caption{Further selection requirements applied to the \Bsl candidate
($m^{PDG}_{D}$ is the nominal $D$ mass~\cite{PDG}).}
\renewcommand\arraystretch{1.2}
\begin{tabular}{lcc}
\hline
\hline
 Variable     & Mode & Range \\
\hline
\cosBY                & \Dz modes  & [-2.00,1.00] \\
	              & \Dpm modes & [-1.00,1.00] \\
\hline
$m_{D} - m^{PDG}_{D}$ & $\Bp \to \Dz(\Km \pip \piz) \ell^+ \nu$ &
[-0.035,0.035] \\
(\gevcc)	      & other $B \to D \ell \nu$ modes &
[-0.020,0.020] \\
\hline
$\Delta m$            & $\Dstarz \to \Dz \gamma$ & [0.10,0.15] \\
(\gevcc)              & $\Dstarpm \to D^{\pm(0)} \pi^{0(\pm)}$    & [0.14,0.15]
\\
\hline
\hline
\end{tabular}
\label{tab:sel_Bsl}
\end{center}
\end{table}

\begin{table*}[t]
\begin{center}
\caption{Expected signal and background yields ($N_s$ and $N_b$ respectively)
from MC studies (assuming the SM $\mathcal{B}$ for the signal) and results of
the data fit, along with signal efficiencies, corrected for systematic effects.
Expected signal yields are evaluated according to the SM expected $\mathcal{B}$.
The first error on the fitted signal yield and on
$N_{\Bhad}$ is statistical, the second is systematic. The corresponding upper
 limits are also quoted.}
  \begin{tabular}{cccc}
 \hline
 \hline
\Kstar mode & ~~~~~~~~~~$K^+ \piz$~~~~~~~~~~ & ~~~~~~~~~~$\KS \pip$~~~~~~~~~~ &
~~~~~~$K^+ \pim$~~~~~~ \\
 \hline
 \multicolumn{4}{c}{SL ANALYSIS} \\
 \hline	
 Expected Yields & & & \\
 $N_{s}$ & 3.31 & 2.54 & 4.07 \\
 $N_{b}$ & 697  & 827  & 468 \\
 \hline
\Eextra Fit Results & & & \\
 $N_{s}$ & -22 $\pm$ 16 $\pm$ 14 & 3 $\pm$ 17 $\pm$ 15 & 35 $\pm$ 13 $\pm$ 9\\
 $N_{b}$ & 754 $\pm$ 32 & 869 $\pm$ 34 & 476 $\pm$ 25 \\
 \hline
 $\varepsilon$ ($\times 10^{-4}$) & 5.6 $\pm$ 0.7 & 4.3 $\pm$ 0.6 & 6.9
$\pm$ 0.8 \\
 $N_{\BB}$ ($\times 10^6$) & \multicolumn{3}{c}{$454 \pm 5$} \\
 \hline 
 UL (90\% CL) & \multicolumn{2}{c}{$9 \times
10^{-5}$} & $18 \times 10^{-5}$  \\
 \hline
 \hline
 \multicolumn{4}{c}{HAD ANALYSIS} \\
 \hline	
 Expected Yields & & & \\
 $N_{s}$ & 0.87 & 0.77 & 1.64\\
 $N_{b}$ & 46 & 35 & 73\\
 \hline
 NN Fit Results & & & \\
 $N_{s}$ & 5 $\pm$ 6 $\pm$ 4 & 3 $\pm$ 7 $\pm$ 4 &
 -10 $\pm$ 9 $\pm$ 6\\
 $N_{b}$ & 39 $\pm$ 9 & 51 $\pm$ 10 & 77 $\pm$ 13\\
 \hline
 $\varepsilon_{\Bsig}$ ($\times 10^{-2}$) & $5.8 \pm 0.5$ & $5.2 \pm 0.6$ &
$16.6 \pm 1.4$ \\
 $N_{\Bhad}$ ($\times 10^5$) & \multicolumn{2}{c} {$10.128 \pm 0.010 \pm 0.344$}
& $7.175 \pm 0.008 \pm 0.222$ \\
 \hline 
 UL (90\% CL) & \multicolumn{2}{c}{$21 \times
10^{-5}$} & $11 \times 10^{-5}$  \\ 
 \hline
\hline
\end{tabular}
\label{tab:results}
\end{center}
\end{table*}

Background contamination is reduced by applying a further selection on the
\BB candidates. Event shape
variables, namely \costhrust\ (the angle between the
tag side reconstructed momentum and the thrust axis~\cite{thrust} of the ROE)
and $R_2$ (the ratio of the second and zeroth Fox-Wolfram moments~\cite{R2}),
are
used to reject the continuum background. The \Kstar mass ($m_{\Kstar}$) and,
for the $\KS\pip$ mode, the \KS mass ($m_{\KS}$) allow rejection of
combinatorial \Kstar candidates. We define the missing 4-momentum due to
unreconstructed neutrinos as the difference between the \FourS
4-momentum and the reconstructed tagging \B and \Kstar 4-momenta. It is
exploited
in the selection through the combination $\Emiss + \pmiss$ (the sum of the
missing energy and the missing momentum evaluated in the CM frame) and the angle
\cosmiss\ (the azimuthal angle of the missing momentum in the CM frame). The
extra neutral energy \Eextra, defined as the sum of the energies of the EMC
neutral clusters not used to reconstruct either the tag or the signal $B$, is
exploited, considering that signal events have no additional neutral particles
produced in association with the \Kstar. The requirements applied on the
selection variables described above are listed in Tab.~\ref{tab:selection}.

In the SL analysis, the selection
is optimized in the MC samples by maximizing the Punzi figure of
merit~\cite{punzi}, given by
$\varepsilon/(n_{\sigma}/2 + \sqrt{N_{b}})$, where
$\varepsilon$ is the total signal efficiency, $N_{b}$ is the number of
expected background data events and $n_{\sigma} = 1.285$
corresponds to a one-side 90\% confidence level. We also refine the \Bsl
selection with respect to the one applied before the choice of the best
candidate, and the corresponding requirements are summarized in
Tab.~\ref{tab:sel_Bsl}, where, $\Delta m$ is the difference between the $D$
and \Dstar masses, expected to be $142.17 \pm 0.07$~\cite{PDG}.
The total signal efficiency, evaluated with MC simulations, is given
in Tab.~\ref{tab:results}. The variable \Eextra is not used in the selection
optimization, and its distribution is used in an extended maximum likelihood fit
in order to extract the signal yield. Due to the lower bound on the
energy of detected photons ($50\,\mev$), the distribution of \Eextra is not
continuous, so we define the likelihood in the following form: 
\begin{widetext}
\begin{eqnarray}
\nonumber \mathcal{L}(N_s,N_b) & = & \frac{e^{- \left[ (1-f_s) N_s + (1-f_b) N_b \right] }}{N_1!} \\
\nonumber & \times & \prod_{i=1}^{N_1} \left[ P_{sig}(E_{extra,i}|\mathbf
p_{sig})(1-f_{s})N_{s} +  P_{bkg}(E_{extra,i}|\mathbf
p_{bkg})(1-f_{b})N_{b} \right] \\
& \times & \frac{(f_s N_s + f_b N_b)^{N_0}e^{- (f_s N_s + f_b N_b)}}{N_0!} \, ,
\label{eq:like}
\end{eqnarray}
\end{widetext}
where $N_s$ and $N_b$ are the expectation values for the numbers
of signal and background events; $f_s$ and $f_b$ are the fractions of signal and
background events with $\Eextra = 0$, and are fixed from the results obtained
in the MC samples; $N_0$ and $N_1$ are the numbers of observed events with
$\Eextra = 0$ and $\Eextra > 50\,\mev$ respectively; and $P_{sig}$ and $P_{bkg}$
are the probability distribution functions (PDF) for signal and background,
depending on a set of parameters
$\mathbf p_{sig}$ and $\mathbf p_{bkg}$ respectively. MC studies show that the
background distribution is well described by a first-order polynomial PDF, while
the signal shape can be parameterized with an exponential function and, in the
charged modes, with an additional Landau contribution that accounts
for photons from a tag side \Dstar not associated to the \Bsl during
the reconstruction. The parameters of the PDFs are evaluated in the MC samples
and fixed when fitting the real data. The fit strategy is validated by means
of simulation studies which do not show any significant bias on the signal
yields.
The fits to the \Eextra distributions in the data sample are shown in
Fig.~\ref{fig:fit} and the fitted yields are quoted in Tab.~\ref{tab:results} 
along with the total efficiencies $\varepsilon$.

In the HAD analysis, we apply a loose selection (Tab.~\ref{tab:selection}), then
all discriminating
variables are used as inputs for a Neural Network (NN), whose output variable
$NN_{\mbox{\scriptsize{out}}}$ is fitted in the region
$NN_{\mbox{\scriptsize{out}}} > 0.6$, 
where the events from the signal MC sample are mostly concentrated. 
The upper bound of the fit region is different among the three \Kstar modes,
reflecting the shape of $NN_{\mbox{\scriptsize{out}}}$ in the signal MC sample.
Three
different NN are trained, one for each \Kstar decay mode. The signal output is
described with an exponential function for the \KstarztoKpi mode and a Crystal
Ball PDF~\cite{CB} for the charged \Kstar channels. The background is
parameterized by
\begin{equation}
 f(x) \propto \frac{x + k_1}{1 + e^{k_2 x}} \, .
\end{equation}
Also in this case, in the fit to real data the signal and background PDF
parameters are fixed to the values extracted from the MC simulations. Simulated
experiments are used to validate the fit strategy. The fits to the
$NN_{\mbox{\scriptsize{out}}}$ distributions in the data sample are
shown in Fig.~\ref{fig:fit}, the fitted yields
and the \Bsig efficiencies $\varepsilon_{\Bsig}$ are quoted in
Tab.~\ref{tab:results}.

\begin{table*}
\begin{center}
\caption{Summary of systematic uncertainties on the signal
efficiency, signal yield, and normalization.}
\begin{tabular}{lcccccc}
\hline
\hline
 & \multicolumn{3}{c}{SL ANALYSIS} & \multicolumn{3}{c}{HAD ANALYSIS}\\
\Kstar mode & ~~~$K^+ \piz~~~$ & ~~~$\KS \pip$~~~ & ~~~$K^+ \pim$~~~ & ~~~$K^+
\piz$~~~ & ~~~$\KS \pip$~~~ & ~~~$K^+ \pi^-$~~~\\
\hline
 \multicolumn{7}{c}{Signal efficiency (\%)} \\
\hline
MC statistics       & 1.4 & 1.7 & 1.3
                    & 2.9 & 3.1 & 2.4 \\
Best pair selection & 0.2 & 0.0 & 0.0 & -- & -- & -- \\
Tagging Efficiency  & 10.0 & 10.0 & 10.0 & -- & -- & -- \\
Tracking            & 0.3 & 1.0 & 0.7
                    & 0.3 & 1.0 & 0.7 \\
\piz reconstruction & 3.0 & -- & --
                    & 3.0 & -- & -- \\
\KS reconstruction  & -- & 2.5 & --
                    & -- & 2.5 & -- \\
Particle  ID          & 1.7 & -- & 1.4 & 1.7 & -- & 1.4 \\
Selection variables & 5.0 & 7.3 & 5.1
                    & 5.3 & 8.6 & 3.8 \\
Model dependence    & 4.5 & 4.8 & 1.3
                    & 6.3 & 7.4 & 6.9 \\
\hline
 \multicolumn{7}{c}{Signal yield (events)} \\
\hline
Signal PDF param.    & 0.7 & 1.4 & 0.2 & 0.2 & 0.3 & 0.2 \\
Bkgd PDF param.      & 11.0  & 11.0  & 7.7 & 2.8 & 2.8 & 4.5 \\
Signal PDF shape    & --   & --   & -- & 1.2 & 1.7 & 1.2\\
Bkgd PDF shape      & 6.4 & 4.9 & 2.8 & 2.1 & 1.6 & 3.4\\
\hline
 \multicolumn{7}{c}{Normalization factor (\%)} \\
\hline
$N_{BB}$ or $N_{\Bhad}$            & 1.1 & 1.1 & 1.1 & 3.4 & 3.4 & 3.1  \\
\hline
\hline
\end{tabular}
\label{tab:syst}
\end{center}
\end{table*}

The branching fraction measurement is affected by systematic uncertainties
related to the signal efficiency estimate, the \BR\ normalization and the signal
yield extraction from the fit.

The signal efficiency has an uncertainty due to the limited MC statistics.
Control samples are used to estimate and correct for data/MC disagreement 
in the charged particle tracking, neutral particle reconstruction and
particle identification.
Uncertainties associated with the event selection criteria are computed
depending on the specific selection strategy. 
Data/MC comparisons and expected detector resolutions
provide an estimate of possible discrepancies in the distribution of the
selection variables. For the SL analysis these values are used to vary the
selection requirements and evaluate the impact on the efficiency;
for the HAD analysis they are used to randomly smear
the distributions of the NN inputs and evaluate the impact on the efficiency
after the NN cut.
The uncertainty due to the residual model dependence of our measurement
is estimated as follows. We apply a weight to each MC event, based on
the generated value of $s_{\nu\nu}$, in such a way that the weighted
distribution matches the expected distribution in the SM or some specific NP
model. Then the efficiency is evaluated taking into account the weights and is
compared to the nominal efficiency (obtained from the unweighted MC events,
generated with a pure phase space model). For the SL analysis two further
uncertainties are associated with the best candidate selection and the \Bsl
selection efficiency. The former is evaluated by modifying the input likelihoods
of Eq.~(\ref{eq:best}) according to data/MC comparisons. Concerning the \Bsl
selection efficiency, we apply a correction given by the square root of the
data/MC ratio of the number of double-tagged events. Alternative approaches are
used to compute the same correction factor and the largest discrepancy with
respect to the nominal approach is taken as a systematic uncertainty.

The error on the number of produced \BB events is estimated 
as described in Ref~\cite{bcount}. The systematic error for
$N_{\Bhad}$, used in the
HAD analysis, is computed by varying the MC \BB component both in shape and
normalization. The ratio 
$\varepsilon^{\mbox{\scriptsize{\Kstar\nunub}}}_{\mbox{\scriptsize{\Bhad}}}
/\varepsilon^{\mbox{\scriptsize{\BB}}}_{\mbox{\scriptsize{\Bhad}}}$,
is used to correct the tag yield and to assign 
further systematic uncertainties. Since the tagging efficiency depends
on the global event multiplicity, this ratio is expected to be different from 1
and to depend on the signal side decay and the \Bhad charge. From MC simulations
it is found to be $1.008 \pm 0.007$ for the charged tag and $1.176 \pm 0.013$
for the neutral one.

The systematic uncertainties associated with the signal yield are due to the
statistical errors on the PDF parameters (fixed from the MC sample) and
potential data/MC disagreement for the shapes. We vary the parameters according
to their statistical error and correlations. The background shapes are validated
in the SL (HAD) analysis with the $m_{D}$ (\mes) sideband: the data/MC ratio of
the fit variable distribution is parameterized by a first-order polynomial, that
is used to modify the nominal background PDF. A similar strategy is adopted
in the SL analysis to validate the signal PDF with double-tagged
events. For the HAD analysis we compare the distributions of the NN output
before and after the smearing of the inputs. In the SL analysis, also the statistical
errors and the data/MC disagreements for the fractions $f_s$ and $f_b$ are included in
these estimates. A summary of the systematic uncertainties is listed in Tab.~\ref{tab:syst}.

No significant signal is observed in the two analyses. A Bayesian approach is
used to set upper limits at the 90\% confidence level on $\BR_\pm =
\BR(\BChtoKChnn)$,
$\BR_0 = \BR(\BNeutoKNeunn)$ and on their combination. Flat prior
probabilities are assumed for positive values of both $\mathcal{B}$'s. Gaussian
likelihoods are adopted for the observed signal yields, related to the
$\mathcal{B}$'s by Eq.~(\ref{eq:BR}) or Eq.~(\ref{eq:BRhad}). The Gaussian
widths are fixed to the sum in quadrature of the statistical and systematic
yield errors. We extract a posterior two-dimensional PDF $P(\BR_\pm,\BR_0)$
using Bayes theorem, including in the calculation the effect of systematic
uncertainties associated with the efficiencies and the normalizations, modeled
by Gaussian PDFs. Systematic uncertainties that are common to the different
channels and to the two analyses are assumed to be fully correlated. The 90\%
confidence level upper limits are calculated, after the marginalization of the
two-dimensional posterior, by:
\begin{equation}
\int_0^{UL}\mathcal{P}_{0,\pm}(\BR_{0,\pm}) \,d\BR_{0,\pm} {\bigg/}
\int_0^{\infty}\mathcal{P}_{0,\pm}(\BR_{0,\pm}) \,d\BR_{0,\pm} = 0.9 \, .
\end{equation}
The cross-feed between the different channels is found to be negligible
in the MC events, but is included in the calculation for completeness. We
extract the combined upper limits:
\begin{eqnarray}
 \nonumber \BR(\BChtoKChnn) &<& 8 \times 10^{-5} \\
 \nonumber \BR(\BNeutoKNeunn) &<& 12 \times 10^{-5} \\
 \BR(\B \to \Kstar \nunub) &<& 8 \times 10^{-5}.
\end{eqnarray}

In summary, we search for \BtoKnn decays in a data sample corresponding
to 413~\invfb, collected by the \babar\ experiment at the \FourS resonance. We
do not observe a significant signal in any of the modes studied and set upper
limits on the decays \BNeutoKNeunn and \BChtoKChnn, and the combined channel
\BtoKnn. Since no constraints were applied to the kinematics of the final state
\Kstar meson, or the undetected \nunub system, these results can be
interpreted in the context of new physics models where invisible particles,
other than neutrinos, are responsible for the missing
energy~\cite{bird,georgi,aliev}. In this way, the results presented here are
model independent. These results represent the most stringent upper limits
reported to date and they are still consistent with the SM
expectation~\cite{BHI}.

We are grateful for the extraordinary contributions of our \pep2\ colleagues in
achieving the excellent luminosity and machine conditions that have made this
work possible. The success of this project also relies critically on the
expertise and dedication of the computing organizations that
support \babar. The collaborating institutions wish to thank
SLAC for its support and the kind hospitality extended to them.
This work is supported by the US Department of Energy and National Science
Foundation, the Natural Sciences and Engineering Research Council (Canada),
the Commissariat \`a l'Energie Atomique and Institut National de Physique
Nucl\'eaire et de Physique des Particules (France), the
Bundesministerium f\"ur Bildung und Forschung and Deutsche
Forschungsgemeinschaft (Germany), the Istituto Nazionale di Fisica Nucleare
(Italy), the Foundation for Fundamental Research on Matter (The Netherlands),
the Research Council of Norway, the Ministry of Education and Science of the
Russian Federation, Ministerio de Educaci\'on y Ciencia (Spain), and the
Science and Technology Facilities Council (United Kingdom).
Individuals have received support from the Marie-Curie IEF program (European
Union) and the A. P. Sloan Foundation.


\begin{thebibliography}{99}
\bibitem{cc}
  Charge conjugation is implied throughout this document,
  unless explicitly stated, and \Kstar refers to $\Kstar(892)$.
\end{thebibliography}
\end{document}